\documentclass[12pt]{article}

\renewcommand{\baselinestretch}{1.7}

\usepackage[dvips]{color}

\usepackage{graphicx}
\usepackage{amssymb}
\usepackage{epsf}
\usepackage{latexsym}

\begin{document}
\begin{titlepage}
\null
\begin{flushright}
%hep-th/YYMMXXX
%HIP-2007-62/TH
\end{flushright}

\vskip 1.2cm

\begin{center}

  {\Large \bf Gauge Covariance of the Aharonov-Bohm Phase \\
  in Noncommutative Quantum Mechanics}

\vskip 2.0cm
\normalsize

  {\large Masud Chaichian\footnote{masud.chaichian@helsinki.fi},
  Miklos L\aa ngvik\footnote{miklos.langvik@helsinki.fi},
  Shin Sasaki\footnote{shin.sasaki@helsinki.fi} and Anca Tureanu\footnote{anca.tureanu@helsinki.fi}}

\vskip 0.5cm

  {\large \it Department of Physics, University of Helsinki, \\
   and Helsinki Institute of Physics \\
              P.O. Box 64, FIN-00014 Helsinki, Finland}

\vskip 1cm

{\bf Abstract}

\end{center}

\renewcommand{\baselinestretch}{1.5}\selectfont

The gauge covariance of the wave function phase factor in noncommutative
quantum mechanics (NCQM) is discussed.
We show that the naive path integral formulation and an approach where one shifts the coordinates
of NCQM in the presence of a background vector potential
leads to the gauge non-covariance of the phase factor.
Due to this fact, the Aharonov-Bohm phase in NCQM which
is evaluated through the path-integral or by shifting the coordinates
is neither gauge invariant nor gauge covariant.
We show that the gauge covariant Aharonov-Bohm effect
should be described by using the noncommutative Wilson lines, what
is consistent with the noncommutative Schr\"odinger equation.
This approach can ultimately be used for deriving an analogue
of the Dirac quantization condition for the magnetic monopole.

\end{titlepage}

\newpage

\section{Introduction}
In the recent decade, there has been a lot of interest in the study of
physics on a noncommutative space-time due to the fact that
space-time may exhibit its noncommutativity at the scale of quantum gravity.
Especially, string theory, which is considered as the most promising
candidate for a theory of quantum gravity, gives rise to space-time
noncommutativity \cite{SeWi}.
Apart from the string theory motivation, it is interesting to investigate
the space-time noncommutativity in a more familiar set-up, like
quantum mechanics. Especially, since the result \cite{Dop}, combining Heisenberg's uncertainty 
principle with Einstein's theory of classical gravity, is quantum mechanical
in spirit, the purely quantum mechanical treatment of a noncommutative space-time becomes interesting.
In \cite{Dop} one considers a gedanken experiment at very high energy where the high
density of the energy-momentum tensor would result in the formation of black holes through 
the Einstein equations. In this case it would no longer be possible
to measure lengths up to arbitrary precision, but space-time would become noncommutative in a
similar way as phase-space becomes noncommutative in quantum mechanics.

Various approaches to quantum mechanics on noncommutative space-time have been proposed in
\cite{Me, DuvHor, ChShTu2, NaPo}. Its space coordinate operator $\hat{X}_i$ is characterized by the relation
\begin{eqnarray}
[\hat{X}_i, \hat{X}_j] = i \theta_{ij}, \label{NC}
\end{eqnarray}
where $i = 1, 2, 3$ stands for the three space coordinates and
the constant $\theta_{ij}$ is the noncommutativity parameter.
Here we have taken the time direction to be commutative $[\hat{X}_0, \hat{X}_i] = 0$, due to the
problems with unitarity \cite{unitarity} and causality \cite{causality} for a noncommuting time direction.
We represent the noncommutativity of space coordinates through the
Weyl-Moyal correspondence, in which to each function of operators
$f(\hat X)$ corresponds a Weyl symbol $f(x)$, defined on the
commutative counterpart of the space. This amounts to replacing the usual commutative
product of functions of operators $f(\hat{X})g(\hat X)$ by the Moyal star-product of
Weyl symbols, $f(x)\star g(x)$, where,
\begin{eqnarray}
(f\star g) (x) = f(x) \exp \left[
\frac{i}{2} \theta_{ij} \overleftarrow{\partial}_i
\overrightarrow{\partial}_j \right] g(x), \label{star}
\end{eqnarray}
and $x$ are the commutative space coordinates.
%and the products among eigenvalues of position operator is evalulated
%in the star product defined above.
The canonical quantization condition
between the quantum mechanical coordinate $\hat{X}_i$ and momentum $\hat{P}_i$
 is the same as in ordinary quantum mechanics;
\begin{eqnarray}
& & [\hat{X}_i, \hat{P}_j] = i \hbar \delta_{ij}, \quad [\hat{P}_i, \hat{P}_j] = 0,
\end{eqnarray}
but with the additional relations
\begin{eqnarray}
[\hat{X}_i, \hat{X}_j] = i \theta_{ij}, \quad
[\hat{X}_i, \theta_{kl}] = [\hat{P}_i, \theta_{kl}] = 0.
\end{eqnarray}
The wave function $\Psi (x)$ now satisfies
\begin{eqnarray}
\hat{P}_i \Psi (x) = - i\hbar \partial_i \Psi (x), \quad
\hat{X}_i \Psi (x) = x_i \star \Psi (x).
\end{eqnarray}
%where $\hat{P}_i$ and $\hat{X}_i$ are the quantum mechanical
%momentum and coordinate operators respectively.
All the wave functions
and any operators which are dependent on the space-time
coordinates should be multiplied by the star product defined above.

In the context of noncommutative quantum mechanics (NCQM),
many observable quantities have been studied. They include the
Aharonov-Bohm (AB) effect \cite{ChDePrShTu2, ChDePrShTu, LiDu}, the hydrogen atom
spectrum and the Lamb shift \cite{ChShTu2, ChShTu}, the Hall effect \cite{DaJe}, the
Aharonov-Casher effect \cite{LiWa} and so on.

Since all the observables in quantum mechanics should be
gauge invariant quantities, it is important to examine the gauge invariance
of physical quantities in NCQM.
For instance, the gauge invariance (or covariance) of the phase factor
of a wave function is directly related to many of the
physical observables, such as, the Aharonov-Bohm effect,
the Aharonov-Casher effect and the Berry phase.

%some citation to the Berry phase?

In this letter, we show that the naive
path integral formulation of NCQM and an approach
where one shifts the coordinates of NCQM \cite{LiDu}
lead neither to a gauge invariant nor to a gauge covariant Aharonov-Bohm phase factor\footnote{The
shift of coordinates of NCQM has previously been used in \cite{Me, ChShTu2, CuFaZa}}.
Instead, we propose a gauge covariant formulation of the AB phase
which is consistent with the noncommutative Schr\"odinger equation.

The organization of this letter is as follows.
In section \ref{PINCQM}, we introduce the path integral formulation of
NCQM following the result of \cite{ChDePrShTu2, ChDePrShTu} especially focusing on
the gauge covariance of the formulation.
We shall stress the difference between the commutative and noncommutative
cases and point out how gauge covariance is broken in the noncommutative case.
Section \ref{Bopp} is devoted to another approach to NCQM where one shifts the coordinates
to satisfy the usual commutation relations of ordinary quantum mechanics. This approach also
breaks gauge invariance but preserves some exotic kind of gauge invariance.
In section \ref{Wilson}, we propose a gauge covariant AB phase factor
which is represented by the path-ordered exponential and is consistent with
the Schr\"odinger equation. Section \ref{DI} contains summary and discussion.

\section{Path integral formulation of NCQM \label{PINCQM}}
In this section, we introduce the path integral formulation
of NCQM following the derivation of \cite{ChDePrShTu2, ChDePrShTu}.
We consider a particle
with mass $m$ and charge $e$,
 under the noncommutative $U(1)$
gauge group, in a magnetic field. The corresponding gauge potential is $A_i \ (i = 1,2,3)$.
In the following, we consider only the case of a time-independent background $A_i (\vec{x})$.
The noncommutative Hamiltonian is given by
\begin{eqnarray}
H(x) = \frac{1}{2m}\Big(P_i + \frac{e}{c}A_i (x) \Big)_{\star}^2, \label{Hamiltonian}
\end{eqnarray}
where $P_i = -i\nabla_i$. The star $U(1)$ gauge field strength is defined by
\begin{eqnarray}
F_{ij} = \partial_{i} A_{j} - \partial_j A_i + i\frac{e}{\hbar c}[A_i, A_j]_{\star}. \label{field_strength}
\end{eqnarray}
The transition amplitude from the initial state $\Psi_i$ to the
final state $\Psi_f$, $(\Psi_f, e^{-\frac{i H t}{\hbar}} \Psi_i)$ is
invariant under the following noncommutative gauge transformations,
\begin{eqnarray}
\Psi (x) & \longrightarrow & U (x) \star \Psi (x), \nonumber \\
A_i (x) & \longrightarrow & U(x) \star A_i (x) \star U^{-1} (x)
- \frac{i\hbar c}{e} U(x) \star \partial_i U^{-1} (x), \nonumber \\
P_i & \longrightarrow & U(x) \star P_i \star U^{-1} (x)
+ i \hbar U(x) \star \partial_i U^{-1} (x). 
\end{eqnarray}
Here $\Psi (x)$ is the wave function and $U(x)$ is defined by
$U(x) = e_{\star}^{-\frac{ie}{\hbar c}\lambda (x)}$ with a real function $\lambda (x)$.
The star $U(1)$ element $U (x)$ satisfies $U^{-1} \star U = U \star U^{-1} = 1$.
The Hamiltonian transforms covariantly under the gauge transformation,
\begin{eqnarray}
H (x) \longrightarrow U (x) \star H (x) \star U^{-1} (x), \label{hamiltonian_gauge_transformation}
\end{eqnarray}
while in the commutative case, $H(x)$ is invariant under the $U(1)$ gauge transformation.
%From this,
%the phase factor that is obtained after the time evolution of state
%should be gauge invariant quantity.

The propagator $\mathcal{K}_t (x,y)$ is represented by the bi-local kernel \cite{ChDePrShTu2, ChDePrShTu}
\begin{equation}
\mathcal{K}_t (x,y) = \langle x | e^{-\frac{i H(x) t}{\hbar}} | y \rangle =
\int \! \frac{d^3 q}{(2 \pi\hbar)^3}
(e^{-\frac{i H (x) t}{\hbar}}
\star
e^{\frac{i q x}{\hbar}}) e^{- \frac{i q y}{\hbar}}.
\end{equation}
Note that the action of $H(x)$ on $e^{iqx\over \hbar}$ is via the 
star-product defined in (\ref{star}).
This propagator is bi-locally gauge covariant provided the
Hamiltonian transforms as in (\ref{hamiltonian_gauge_transformation}).
The naive gauge transformation of $\mathcal{K}_t (x,y)$ is
explicitly given by
\begin{eqnarray}
\mathcal{K}_t (x,y) \longrightarrow \mathcal{K}'_t (x,y)
&=& \int \! \frac{d^3 q}{(2\pi\hbar)^3} \left(
U(x) \star_x e^{-\frac{i H(x) t}{\hbar}} \star_x U^{-1} (x) \star_x
U(x) \star_{x} e^{\frac{i qx}{\hbar}} \right)
\left(
e^{- \frac{i qy}{\hbar}} \star_y U^{-1} (y)
\right) \nonumber \\
&=& U(x) \star_x \mathcal{K}_t (x,y) \star_y U(y)^{-1},
\end{eqnarray}
where we have used the gauge transformations 
\begin{eqnarray}
H(x) & \rightarrow & U(x)\star_xH(x)\star_xU^{-1}(x) \nonumber \\
e^{iqx\over \hbar} & \rightarrow & U(x)\star_x e^{iqx\over \hbar} \nonumber \\
e^{-{iqy\over \hbar}} & \rightarrow & e^{-{iqy\over \hbar}}\star_yU^{-1}(y). \nonumber
\end{eqnarray}
Here $\star_x, \ \star_y$ are the star products defined with respect to $x_i$ and $y_i$,
respectively.
This bi-local covariance guarantees the gauge invariance of the
probabilities, % and also the gauge covariance of the AB phase.
and should provide the gauge covariant AB phase in the path-integral
formulation of NCQM.

The propagator can be represented by the products of short-time
propagators in the infinite time evolution by
separating the time interval into $N$-pieces and
taking $N \to \infty$,
\begin{eqnarray}
\mathcal{K}_t (x,y) = \lim_{N \to \infty} \int \!
d^3 x_{N-1} \cdots d^3 x_1 \
\mathcal{K}_{\epsilon} (x, x_{N-1}) \cdots
\mathcal{K}_{\epsilon} (x_2, x_1) \mathcal{K}_{\epsilon} (x_1, y).
\end{eqnarray}
Here $\epsilon \equiv t / N$ and we have used the identity
$e^{-\frac{i H t_1}{\hbar}} e^{-\frac{i H t_2}{\hbar}} = e^{- \frac{i H}{\hbar}(t_1 + t_2)}$.
The reason why gauge covariance is lost in \cite{ChDePrShTu2, ChDePrShTu} is that the quantum mechanical Hamiltonian
corresponding to (\ref{Hamiltonian}) should be treated in the
Weyl ordered form if we use the midpoint prescription in the path-integral formulation.
This in turn is a consequence of that
the Hamiltonian contains a mixing term between
$\hat{P}_i$ and $\hat{X}_i$.
%This Weyl ordered quantum mechanical
%Hamiltonian leads to the midpoint prescription
%in the path-integral formulation.
This means that the short-time
propagator has to be evaluated in the midpoint of $x$ and $y$,
and we must use $H (\bar{x})$ where $\bar{x}_i = (x_i + y_i)/2$ instead of $H(x)$.
In this case, the propagator is not bi-locally gauge covariant
anymore\footnote{There is another problem with the midpoint prescription in NCQM.
There is an ambiguity in how to define the star product
between $e^{-i \frac{H(\bar{x})t}{\hbar}}$ and $e^{\frac{iqx}{\hbar}}$ in the kernel.
Here we have simply assumed that it is given by $\star_{\bar{x}}$. It could also be given
by $\star_{x}$, but this does not change the outcome. The propagator is still not bi-locally gauge covariant.},
\begin{eqnarray}
\mathcal{K}_t (x,y) \longrightarrow \mathcal{K}'_t (x,y)
&=& \int \! \frac{d^3 q}{(2\pi\hbar)^3} \left(
U(\bar{x}) \star_{\bar{x}} e^{-\frac{i H(\bar{x}) t}{\hbar}} \star_{\bar{x}} U^{-1} (\bar{x}) \star_{\bar{x}}
U(x) \star_{x} e^{\frac{i qx}{\hbar}} \right)
\left(
e^{- \frac{i qy}{\hbar}} \star_y U^{-1} (y)
\right) \nonumber \\
&\not=& U(x) \star_x \mathcal{K}_t (x,y) \star_y U(y)^{-1}.
\end{eqnarray}
We would like to stress that the propagator is bi-locally gauge covariant
in the commutative case, namely,
\begin{eqnarray}
\mathcal{K}_t (x,y) \longrightarrow \mathcal{K}'_t (x,y)
= U(x) \mathcal{K}_t (x,y) U^{-1} (y) \quad \textrm{(commutative case).}
\end{eqnarray}
If one goes ahead with the midpoint prescription in the noncommutative case, one arrives at a phase shift
$\delta \phi$ for an electron wave function after moving around the path $C$ in the noncommutative space given by
\begin{eqnarray}
\delta \phi = \frac{e}{\hbar c}\oint_C d x_i \ A_i
+ \frac{em}{4\hbar^2 c} \vec{\theta} \cdot \int_C d x_i \
\left[ (\vec{v} \times \vec{\nabla} A_i) - \frac{e}{mc} (\vec{A} \times
\vec{\nabla} A_i ) \right] + \mathcal{O} (\theta^2). \label{phase}
\end{eqnarray}
Here the component of $\vec{\theta}$ is defined by $\theta_{i} =
\varepsilon_{ijk} \theta_{jk}$.
This is the result obtained in the path-integral formulation in the midpoint prescription
\cite{ChDePrShTu2, ChDePrShTu}. The same result has been obtained by the
perturbative analysis of the Schr\"odinger equation \cite{FaGaLoMeRo}.

We can explicitly check that this result is neither
gauge invariant nor covariant under the $\mathcal{O} (\theta)$ gauge
transformations
\begin{eqnarray}
& & \delta A_i^{(0)} = - \partial_i \lambda, \nonumber \\
& & \delta A_i^{(1)} = \frac{e}{\hbar c} \theta_{kl} \partial_k A_i
\partial_l \lambda.
\end{eqnarray}
Here $A_i^{(n)}$ is an $n$-th order expansion of $A_i$ in the noncommutativity
parameter $\theta_{ij}$.
As we mentioned, this gauge non-covariance originates from the Weyl ordering
 of the quantum mechanical Hamiltonian and hence, from the midpoint prescription in the path-integral.
In the next section, we will use another approach to derive the the AB phase in NCQM.
From here on, for simplicity, we shall use $\hbar = c = m = e = 1$.

\section{The phase shift in terms of a shift of coordinates \label{Bopp}}
It is known that the noncommutativity of space in quantum mechanics
can be interpreted as ordinary quantum mechanics with
deformed Hamiltonian. This deformation can be performed via a
shift of coordinates \cite{Me, ChShTu2, CuFaZa}.

Consider quantum mechanics
on a noncommutative space, with the commutation relation among
coordinate and momentum operators as
\begin{eqnarray}
[\hat{X}_i, \hat{X}_j] = i \theta_{ij}, \quad [\hat{X}_i, \hat{P}_j] = i \delta_{ij},
\quad [\hat{P}_i, \hat{P}_j] = 0.
\end{eqnarray}
Following the procedure adopted in \cite{Me, ChShTu2}, the shifted coordinate and
momentum
\begin{eqnarray}
\hat{x}_i & = & \hat{X}_i + \frac{1}{2} \theta_{ij} \hat{P}_j \\
\hat{p}_i  & = & \hat{P}_i,
\end{eqnarray}
satisfy
\begin{eqnarray}
[\hat{x}_i, \hat{x}_j] = 0, \quad [\hat{x}_i, \hat{p}_j] = i
\delta_{ij}, \quad [\hat{p}_i, \hat{p}_j] = 0.
\end{eqnarray}
Thus NCQM now reduces to ordinary quantum mechanics but
with deformed Hamiltonian $H (\hat{X}, \hat{P}) \to \tilde{H} (\hat{x}, \hat{p})$.
The gauge potential in the Hamiltonian can be expanded as
\begin{eqnarray}
A_i (\hat{X}) = A_i (\hat{x}) - \frac{1}{2} \theta_{kl} \hat{p}_l \partial_k
 A_i(\hat{x}) + \mathcal{O} (\theta^2).
\end{eqnarray}
Consequently, the noncommutative Hamiltonian $H(\hat{X}, \hat{P}) = \frac{1}{2} (\hat{P}_i + A_i
(\hat{X}) )^2$ is interpreted as the deformed Hamiltonian
\begin{eqnarray}
\tilde{H}(\hat{x}, \hat{p}) = \frac{1}{2} \left( \hat{p}_i - A_i (\hat{x}) - \frac{1}{2} \theta^{kl} \hat{p}_l \partial_k
A_i (\hat{x}) \right)^2 + \mathcal{O} (\theta^2),
\label{ham}
\end{eqnarray}
in ordinary quantum mechanics. The Hamiltonian (\ref{ham}) is no longer star-gauge
covariant as a consequence of shifting the coordinates. This is because
the potential $A_i(\hat{X})$ is given in the noncommutative space and it transforms
as
\begin{equation}
A_i(\hat{X}) \rightarrow A'_i(\hat{X}) = U(\hat{X}) A_i(\hat{X}) U^{-1}(\hat{X}) - iU(\hat{X})\partial_iU^{-1}(\hat{X}).
\label{atrans}
\end{equation}
However, the potential $A_i(\hat{x})$ is not given in this type of noncommutative space, but the ordinary
quantum mechanical one, and consequently does not transform similarly to (\ref{atrans}). Therefore,
the star gauge covariance of the Hamiltonian is lost in (\ref{ham}).

The Schr\"odinger equation corresponding to (\ref{ham}) is
\begin{eqnarray}
i \frac{\partial}{\partial t} \Psi(\hat{x})
= \frac{1}{2} \left( \hat{p}_i - A_i (\hat{x}) - \frac{1}{2} \theta_{kl} \hat{p}_l \partial_k
A_i (\hat{x}) \right)^2 \Psi(\hat{x}).
\label{boppschro}
\end{eqnarray}
The solution to this equation is obtained from the commutative solution
through the shift of coordinates
\begin{eqnarray}
\Psi (x) = \psi (x) \exp \left[ i \int^x \! d \xi_i \ (A_i (\xi) + \frac{1}{2} \theta_{kl}
p_l \partial_j A_i (\xi)) \right],
\end{eqnarray}
where $\psi$ is the solution of the equation with vanishing gauge potential and $p_l$ is now the 
eigenvalue of $\hat{p}_l$  as $\hat{p}_l$ only acts on $\Psi$ in (\ref{boppschro}) because of the
antisymmetry of $\theta_{kl}$.
It was shown  \cite{LiDu} that the phase shift in this solution is equivalent to the path
integral result obtained in \cite{ChDePrShTu2, ChDePrShTu}, {\it i.e.} equation
(\ref{phase}) in the previous section and thus is neither gauge invariant
nor covariant.

A comment is in order about the gauge invariance of this approach.
In view of the shifted coordinate, the Hamiltonian and
any physical observables are manifestly invariant under the
coordinate shifted gauge transformation but not under the
ordinary star gauge transformation.
Here the coordinate shifted gauge transformation is defined by the
commutative $U(1)$ gauge transformation evaluated in the shifted
coordinate $x_i - \frac{1}{2} \theta_{ij} p_j$.

\section{The gauge covariant phase factor: the Wilson loop \label{Wilson}}
In this section we propose a gauge covariant phase factor
which can be obtained with the help of the Wilson loop operator.
Let us first consider the AB phase in commutative quantum mechanics.
The Schr\"odinger equation in the presence of a time independent vector potential is
\begin{eqnarray}
i \frac{\partial}{\partial t} \Psi_{\mathrm{Comm}} =
\frac{1}{2} \left(
p_i + A_i (x) \right)^2 \Psi_{\mathrm{Comm}}.
\end{eqnarray}
This equation is solved by
\begin{eqnarray}
\Psi_{\mathrm{Comm}} (x, t) = \psi (x, t) \exp \left[ -i \int^{x}_C d \xi_i \ A_i
(\xi) \right]. \label{commcase}
\end{eqnarray}
Here $\psi(x, t)$ is the solution of the Schr\"odinger equation in the absence of
the vector potential. The integral is performed along a path $C$ which
ends in the point $x$.

The phase factor $\exp\big[-i \int_C^x d\xi_iA_i(\xi)\big]$ in
(\ref{commcase}) is clearly gauge invariant under the $U(1)$ gauge
transformation $\delta A_i = - \partial_i \lambda (x)$. The AB phase
in the commutative case is evaluated as the gauge invariant magnetic
field $\vec{B}$ through Stokes theorem $\oint_C
d\vec{\xi}\cdot\vec{A} = \int_S d\vec{S}\cdot\vec{B}$ where the
boundary of $S$ is the closed path $C$. Consequently the observable
is gauge invariant (see, e.g., \cite{merz,pascual}).

On the other hand, the Schr\"odinger equation in NCQM is
\begin{eqnarray}
i \frac{\partial}{\partial t} \Psi(x, t) =
\frac{1}{2} \left( \hat{P}_i + A_i (x) \right)^2 \star_x \Psi(x, t), \label{NC_schrodinger}
\end{eqnarray}
where all $x$-dependent terms are evaluated by the star product with respect to $x$. We
recall that a gauge invariant quantity in a non-Abelian gauge
theory is the Wilson loop. Wilson loops have been previously used in
the context of noncommutative gauge field theories for constructing
observable quantities, as well as new representations of the
noncommutative gauge groups, forbidden by the no-go theorem of
noncommutative gauge theories (see e.g. \cite{GrHaIt, ChuDorKhoTra,
ChKoTuArSaUe} and references therein). They are defined by the gauge
trace of the path-ordered exponential. Inspired by this, we consider
the Ansatz for the solution to (\ref{NC_schrodinger}) as
\begin{eqnarray}
\Psi (x,x_0,t) = \mathrm{P} \exp_{\star_{x_0}}\left[ - i \int^{1}_{0} \!
ds{d\xi_i\over ds} \ A_i (x_0 + \xi(s) ) \right] \star_{x_0} \psi (x,x_0,t). \label{Ansatz}
\end{eqnarray}
Here the symbol P stands for path ordering. The parameter $0 \le s
\le 1$ parametrizes the path $C$ with endpoints $x_0 + \xi (0) = x_0$ and
$x_0 + \xi(1) = x_0 + l = x$, where $\xi(0) = 0$ and $\xi(1) = l$. $\psi(x,x_0,t)$ is the
solution of the free Schr\"odinger equation
\begin{equation}
-\nabla^2_x\psi(x,x_0,t) = i{\partial\psi(x,x_0,t)\over \partial t}.
\end{equation}
In the case of the AB experiment, $x_0$ represents the location of
the source of electrons and $x$ represents the point at which the
intensity of the beam is evaluated. The free solution
$\psi(x,x_0,t)$ can also be viewed as a wavefunction at the point
$(x_0, t_0)$ from which it is taken to $(x,t)$ by the free
propagator, ${\cal K}_{\mathrm{free}}(x,t;x_0,t_0)$.

The definition of the path-ordered exponential is
\begin{eqnarray}
\mathcal{U} (x, x_0, C) &\equiv& \mathrm{P} \exp_{\star_{x_0}}\left[ - i \int^{1}_{0} \!
ds{d\xi_i\over ds} \ A_i (x_0 + \xi(s) ) \right] \nonumber \\
&=& 1 + \sum^{\infty}_{n=1} (-i)^n \int_0^1 \! d s_1 \ \int^{s_1}_0 \! d s_2
\cdots \int^{s_{n-1}}_0 \! d s_n \
\frac{d \xi_{i_1} (s_1)}{d s_1} \cdots \frac{d \xi_{i_n} (s_n)}{d s_n}
\nonumber \\
& & \qquad \qquad \times A_{i_1} (x_0 + \xi (s_1)) \star_{x_0} \cdots \star_{x_0} A_{i_n} (x_0 + \xi (s_n)).
\label{po-exp}
\end{eqnarray}
This is nothing but a Wilson line in noncommutative gauge theory \cite{GrHaIt} and under NC gauge transformations
it transforms as:
\begin{eqnarray}\label{wilson_line}
\mathcal{U} (x, x_0, C) \longrightarrow U(x) \star_x \mathcal{U} (x, x_0,
C) \star_{x_0} U^{-1} (x_0). \label{gauwiltra}
\end{eqnarray}
It can be shown (see Appendix A) that this path ordered exponential
satisfies the equation
\begin{eqnarray}
\vec{\nabla}_x \mathcal{U} (x, x_0, C) = - i \vec{A}(x) \star_x \mathcal{U} (x, x_0, C).
\label{Ansatz2}
\end{eqnarray}
%and transforms under a gauge transformation as:
%\begin{eqnarray}\label{wilson_line}
%\mathcal{U} (x, x_0, C) \longrightarrow U(x) \star_x \mathcal{U} (x, x_0,
%C) \star_{x_0} U^{-1} (x_0).
%\end{eqnarray}
%This gauge transformation, together with eqn. (\ref{psi0trans}), guarantees that the
%solution of the NC Schr\"odinger equation (\ref{NC_schrodinger}) transforms as
%\begin{equation}
%\Psi(x,x_0,t) \rightarrow U(x)\star_x\Psi(x,x_0,t),
%\end{equation}
%under a gauge transformation and that the NC Schr\"odinger equation (\ref{NC_schrodinger})
%transforms covariantly with the solution (\ref{Ansatz}) as it should.

Let us check the Ansatz (\ref{Ansatz}), starting with the r.h.s. of
the NC Schr\"odinger equation (\ref{NC_schrodinger}), which reads:
\begin{eqnarray}
H \star_x \Psi = \frac{1}{2} \left[
- \vec{\nabla}^2 \Psi - 2i \vec{A} \star_x
\vec{\nabla} \Psi - i (\vec{\nabla} \cdot \vec{A})
\star_x \Psi + \vec{A} \star_x \vec{A} \star_x \Psi
\right]. \label{NCschro}
\end{eqnarray}
For the evaluation of (\ref{NCschro}) we shall need:
\begin{eqnarray}
\vec{\nabla} \Psi & = & - i \vec{A} \star_x e_P \star_{x_0} \psi
+ e_P \star_{x_0} \vec{\nabla} \psi \\
\vec{\nabla}^2 \Psi &=& - i (\vec{\nabla} \cdot \vec{A})
\star_x e_P \star_{x_0} \psi + i^2 \vec{A} \star_x
\vec{A} \star_x e_P \star_{x_0} \psi \nonumber \\
& & - i \vec{A} \star_x e_P \star_{x_0} \vec{\nabla} \psi
- i \vec{A} \star_x e_P \star_{x_0} \vec{\nabla} \psi + e_P
\star_{x_0} \vec{\nabla}^2 \psi,
\end{eqnarray}
where $\Psi = e_P\star_{x_0}\psi$ and $e_P$ stands for $\mathrm{P} \exp_{\star_{x_0}}\left[ - i \int^{1}_{0} \!
ds{d\xi_i\over ds} \ A_i (x_0 + \xi(s) ) \right]$.

The l.h.s. of the NC Schr\"odinger equation (\ref{NC_schrodinger}) is
\begin{eqnarray}
i \frac{\partial}{\partial t} \Psi
&=& e_P \star_{x_0} i \frac{\partial}{\partial t} \psi \nonumber \\
&=& - \frac{1}{2} e_P \star_{x_0} \vec{\nabla}^2 \psi \nonumber \\
&=& - \frac{1}{2} \left[
\vec{\nabla}^2 \Psi + i (\vec{\nabla} \cdot \vec{A})
\star_x e_P \star_{x_0} \psi + \vec{A} \star_x \vec{A}
\star_x e_P \star_{x_0} \psi + 2 i \vec{A} \star_x e_P \star_{x_0}
\vec{\nabla} \psi \right] \nonumber \\
&=& \frac{1}{2} \left[ - \vec{\nabla}^2 \Psi - i (\vec{\nabla} \cdot \vec{A}) \star_x \Psi
- 2 i \left\{ \vec{A} \star_x e_P \star_{x_0} \vec{\nabla}
\psi - i \vec{A} \star_x \vec{A} \star_x e_P \star_{x_0} \psi \right\}
+  \vec{A} \star_x \vec{A} \star_x \Psi \right] \nonumber \\
&=& \frac{1}{2} \left[ - \vec{\nabla}^2 \Psi
- i (\vec{\nabla} \cdot \vec{A}) \star_x \Psi
- 2i \vec{A} \star_x \vec{\nabla} \Psi
+ \vec{A} \star_x \vec{A} \star_x \Psi \right].
\end{eqnarray}
This is exactly $H \star_x \Psi$ as in (\ref{NCschro}). Thus the Ansatz (\ref{Ansatz}) satisfies
\begin{eqnarray}
i \frac{\partial}{\partial t} \Psi = H \star_x \Psi.
\label{NCschrod}
\end{eqnarray}

\begin{figure}[htb]
\begin{center}
\includegraphics[scale=.6]{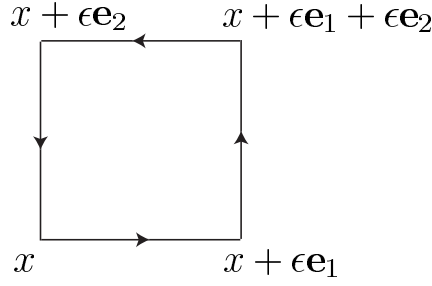}
\end{center}
\caption{Closed path in 1-2 plane}
\label{closed_path}
\end{figure}

The path ordered exponential (\ref{po-exp}) is hard to evaluate explicitly but it can be done
for an infinitesimal closed path $C_l$ in the 1-2 plane depicted in fig.\ref{closed_path}. We can show that
%%%%%%%%%%%%%%%%%%%%%%%%%%%%%%%%%%%%%%%%%%%%%%%%%%%%%%%%%%%%%%%%%%%%%
%%%%%%%%%%%%%%%%%%%%%%%%%%%%%%%%%%%%%%%%%%%%%%%%%%%%%%%%%%%%%%%%%%%%%

\begin{eqnarray}
\mathcal{U} (x, x, C_l) &\equiv&
\mathcal{U} (x, x + \epsilon \mathbf{e_2})
\star_x
\mathcal{U} (x + \epsilon \mathbf{e_2}, x + \epsilon \mathbf{e_1} + \epsilon \mathbf{e_2})
\star_x
\mathcal{U} (x + \epsilon \mathbf{e_1} + \epsilon \mathbf{e_2}, x + \epsilon \mathbf{e_1})
\star_x
\mathcal{U} (x + \epsilon \mathbf{e_1}, x) \nonumber \\
&=& \exp_{\star_x} \left[ \frac{}{} - i \epsilon^2 \left( \partial_1 A_2 (x)
- \partial_2 A_1 (x) \right) + \epsilon^2 [A_1 (x), A_2 (x)]_{\star_x} \right] + \mathcal{O} (\epsilon^3)
\nonumber \\
&=& \exp_{\star_x} \left[ - i \epsilon^2 F_{12} \right] + \mathcal{O} (\epsilon^3),
\end{eqnarray}
where $\epsilon \ll 1$ is the infinitesimal parameter and $\mathbf{e_1},
\mathbf{e_2}$ are unit vectors along the directions 1 and 2. The star product is
evaluated at $x$ and the field strength is defined by (\ref{field_strength}).
The result is manifestly gauge covariant. A generalization of this result to
$U_{\star}(N)$ is possible by replacing $A_i$ by $A_i^aT^a$, where $T^a$ are the
generators of $U(N)$.

The NCAB phase factor for a path $a$ from $x_0$ to $x$ is given by
\begin{equation}
e^{i\delta\phi_{NC}(x, x_0, a)} = \mathrm{P} \exp_{\star_{x_0}}\left[ - i \int^{1}_{0} \!
ds{d\xi_i\over ds} \ A_i (x_0 + \xi(s) ) \right],
\end{equation}
where the path $a$ is parametrized appropriately in the line integral. In view of the
gauge transformation (\ref{gauwiltra}), it transforms as
\begin{equation}
e^{i\delta\phi_{NC}(x,x_0,a)}  \rightarrow  U(x)\star_x
e^{i\delta\phi_{NC}(x,x_0,a)}\star_{x_0} U^{-1}(x_0),
\label{gauwil}
\end{equation}
under a gauge transformation.

The path-ordered phase factor appearing here is
quite similar to the non-Abelian counterpart of the AB phase \cite{WuYa}.
This would be related to the topological features of the phase factor
which will be studied elsewhere \cite{ChLaSaTu}.

One important consistency check for the Ansatz (\ref{Ansatz}) is its
gauge covariance. The wave function $\Psi(x,x_0,t)$ has to transform
in the fundamental representation of $U_\star(1)$, and its Hermitian
conjugate, correspondingly, in the antifundamental representation,
\begin{eqnarray}\label{gauge_tr_sol}
\Psi(x,x_0,t) & \rightarrow & U(x)\star_x\Psi(x,x_0,t), \cr
\Psi^{\dagger}(x,x_0,t) & \rightarrow &
\Psi^{\dagger}(x,x_0,t)\star_x U^{-1}(x)\,,
\end{eqnarray}
in order to insure the gauge covariance of the NC Schr\"odinger
equation. One can show that the gauge transformation (\ref{gauwiltra})
of the path ordered exponential is compatible with this gauge
covariance requirement. Indeed, since $\Psi(x,x_0,t)$ is a solution of
the NC Schr\"odinger equation (\ref{NC_schrodinger}) with the
initial condition $\Psi(x,x_0,t_0)= \Psi(x_0,t_0)$, it follows that,
according to (\ref{gauge_tr_sol}), the initial condition will
transform under gauge transformations as
\begin{equation}\label{gauge_tr_init}
\Psi(x_0,t_0)  \rightarrow  U(x_0)\star_{x_0}\Psi(x_0,t_0)\,.
\end{equation}
On the other hand, the formal general solution of
(\ref{NC_schrodinger}) can be written using the total propagator
${\cal K}(x,t;x_0,t_0)$:
\begin{equation}\label{sol_propagator}
\Psi(x,x_0,t)={\cal K}(x,t;x_0,t_0)\star_{x_0}\Psi(x_0,t_0)\,.
\end{equation}
The total propagator factorizes into the free propagator and the
gauge-field-dependent phase factor, such that the solution can be
written as:
\begin{eqnarray}
\Psi (x,x_0,t) = \mathrm{P} \exp_{\star_{x_0}}\left[ - i
\int^{1}_{0} \! ds{d\xi_i\over ds} \ A_i (x_0 + \xi(s) ) \right]
\star_{x_0} {\cal
K}_{\mathrm{free}}(x,t;x_0,t_0)\star_{x_0}\Psi(x_0,t_0)\,.
\label{Ansatz3}
\end{eqnarray}
By comparing (\ref{Ansatz}) with (\ref{Ansatz3}), it is clear that
\begin{equation}
\psi(x,x_0,t) = {\cal
K}_{\mathrm{free}}(x,t;x_0,t_0)\star_{x_0}\Psi(x_0,t_0)\,,
\end{equation}
and, in view of the fact that the free propagator does not transform
under gauge transformations, while the initial solution
$\Psi(x_0,t_0)$ transforms as (\ref{gauge_tr_init}), the solution
$\psi(x,x_0,t)$ of the free Schr\"odinger equation will have the
peculiar gauge transformation:
\begin{equation}
\psi(x,x_0,t) \rightarrow U(x_0)\star_{x_0}\psi(x,x_0,t)\,.
\label{psi0trans}
\end{equation}
We should emphasize out that $\psi(x,x_0,t)$ is not actually a genuine
solution of a free Schr\"odinger equation, but an artifact of the
factorization of the total propagator as in (\ref{Ansatz3}). In
other words, from the dynamical point of view $\psi(x,x_0,t)$
satisfies the free Schr\"odinger equation, while inheriting at the
same time the gauge transformation property (\ref{gauge_tr_init}) of
the initial solution of (\ref{NC_schrodinger}).

The gauge transformations (\ref{gauwiltra}) and (\ref{psi0trans})
provide the consistency check for the gauge covariance of
$\Psi(x,x_0,t)$ defined by the Ansatz (\ref{Ansatz}). As a result,
the noncommutative Schr\"odinger equation (\ref{NC_schrodinger}) is
covariant under a noncommutative gauge transformation. This
guarantees that the observable probability density $P(x,x_0,t)$, for
the AB-effect of two waves differing by a phase depending on the
paths $a$ or $b$,
\begin{eqnarray}
P(x,x_0,t) & = & \left(\psi^\dagger(x,x_0,t)\star_{x_0} e^{-i\delta\phi_{NC}(x,x_0,a)} +
\psi^\dagger(x,x_0,t)\star_{x_0} e^{-i\delta\phi_{NC}(x,x_0,b)}\right)\star_x \nonumber \\
& & \left(e^{i\delta\phi_{NC}(x,x_0,a)}\star_{x_0}\psi(x,x_0,t) + e^{i\delta\phi_{NC}(x,x_0,b)}\star_{x_0}\psi(x,x_0,t)\right),
\end{eqnarray}
is gauge invariant.

\section{Summary and discussion \label{DI}}
In this letter, we have studied the gauge covariance of the
wave function phase factor in the framework of NCQM.

Due to the fact that the phase factor in a wave function
is frequently related to a physical observable,
it is important to investigate the gauge invariance and covariance of it
in NCQM. The AB phase factor is probably the most
familiar observable phase factor in quantum mechanics.

The naive path-integral formulation of NCQM violates
the star gauge covariance of the AB phase.
The origin of this violation comes from the Weyl ordered quantum mechanical
Hamiltonian and midpoint prescription in the short-time propagator.
This is quite different from the commutative case where
the Hamiltonian itself is $U(1)$ gauge invariant and hence
the propagator is bi-locally gauge covariant.

The same result is obtained by shifting the coordinates of NCQM, whence
the star $U(1)$ gauge invariance/covariance is broken.
However, some exotic gauge invariance, the ''shifted gauge
invariance'' (See end of section 3) is preserved although
the physical meaning of this type of gauge invariance is not clear.

We have found a gauge covariant AB phase factor which is defined
by the path-ordered exponential. This resembles the well-known Wilson
loop in non-Abelian gauge theory.
We have shown that the path-ordered exponential is consistent with the
noncommutative Schr\"odinger equation.
We would like to stress that our result is quite similar to
the non-Abelian AB phase proposed in \cite{WuYa}.
This is very natural because the star $U(1)$ gauge symmetry
is essentially non-Abelian, which can be seen from eq. (\ref{field_strength}).

The AB phase factor is related to the Dirac monopole quantization
and topological properties of the theory and it would be
interesting to find the gauge invariant quantization condition
corresponding to the noncommutative Dirac monopole, especially due
to the results in \cite{HaHaGrNeCiSch} on noncommutative monopoles, dyons and
solitonic solutions. It would also be interesting to investigate the star
gauge invariant path-integral formulation of NCQM \cite{ChLaSaTu}.
\\
\\
%%%%% Acknowledgements  %%%%%%
\noindent {\large \bf Acknowledgments}
\\
We are indebted to Masato Arai, Peter Pre\v{s}najder and Sami Saxell for discussions and useful comments.
The work of S.~S. is supported by the bilateral program of Japan Society
for the Promotion of Science (JSPS) and Academy of Finland, ``Scientist
Exchanges.'' A.~T. acknowledges the grant no. 121720 of the Academy of Finland.

%%%%% Appendix %%%%%

\section*{Appendix A}
\makeatletter
\def\@eqnnum{{\normalfont \normalcolor (\Alph{section}.\theequation)}}
\makeatother
\setcounter{section}{1}
\setcounter{equation}{0}

In this appendix the relation
\begin{equation}
{d\over dx_i} \mathcal{U} (x, x_0, C) = - i A_i(x) \star_x \mathcal{U} (x, x_0, C),
\label{wilrel}
\end{equation}
where
\begin{eqnarray}
\mathcal{U} (x, x_0, C) &\equiv& \mathrm{P} \exp_{\star_{x_0}}\left[ - i \int^{1}_{0} \!
ds{d\xi_i\over ds} \ A_i (x_0 + \xi(s) ) \right] \nonumber \\
&=& 1 + \sum^{\infty}_{n=1} (-i)^n \int_0^1 \! d s_1 \ \int^{s_1}_0 \! d s_2
\cdots \int^{s_{n-1}}_0 \! d s_n \
\frac{d \xi_{i_1} (s_1)}{d s_1} \cdots \frac{d \xi_{i_n} (s_n)}{d s_n}
\nonumber \\
& & \qquad \qquad \times A_{i_1} (x_0 + \xi (s_1)) \star_{x_0} \cdots \star_{x_0} A_{i_n} (x_0 + \xi (s_n)),
\label{po-exp2}
\end{eqnarray}
is proven. The parametrization of the path $C$ is as follows: $x = x_0 + \xi(1) = x_0 + l$ and $x_0 = x_0 + \xi(0)$, so that
$\xi(1) = l$ and $\xi(0) = 0$.

We will begin by considering the path ordered exponential as a continuous function of the parameter $s'$ in the form
\begin{eqnarray}
\mathcal{U} (x(s'), x_0, C) & = &  1 + \sum^{\infty}_{n=1} (-i)^n \int_0^{s'} \! d s_1 \ \int^{s_1}_0 \! d s_2
\cdots \int^{s_{n-1}}_0 \! d s_n \
\frac{d \xi_{i_1} (s_1)}{d s_1} \cdots \frac{d \xi_{i_n} (s_n)}{d s_n}
\nonumber \\
& & \qquad \qquad \times A_{i_1} (x_0 + \xi (s_1)) \star_{x_0} \cdots \star_{x_0} A_{i_n} (x_0 + \xi (s_n)).
\end{eqnarray}
This can be differentiated with respect to $s'$ using the result
\begin{equation}
\partial_b\int_a^b f(x)dx = f(b).
\end{equation}
It gives
\begin{eqnarray}
\partial_{s'} \mathcal{U} (x(s'), x_0, C) & = & \partial_{s'} \sum^{\infty}_{n=1} (-i)^n \int_0^{s'} \! d s_1 \ \int^{s_1}_0 \! d s_2
\cdots \int^{s_{n-1}}_0 \! d s_n \
\frac{d \xi_{i_1} (s_1)}{d s_1} \cdots \frac{d \xi_{i_n} (s_n)}{d s_n}
\nonumber \\
& &\times A_{i_1} (x_0 + \xi (s_1)) \star_{x_0} \cdots \star_{x_0} A_{i_n} (x_0 + \xi (s_n)) \\
& = &  -i\frac{d \xi_{i_1} (s')}{d s'}A_{i_1} (x_0 + \xi (s')) \star_{x_0} \Big[1 + \sum^{\infty}_{n=2}(-i)^{n-1} \ \int^{s'}_0 \! d s_2
\cdots \int^{s_{n-1}}_0 \! d s_n \nonumber \\
& & \times \frac{d \xi_{i_2} (s_2)}{d s_2} \cdots \frac{d \xi_{i_n} (s_n)}{d s_n}
A_{i_2} (x_0 + \xi (s_2)) \star_{x_0} \cdots \star_{x_0} A_{i_n} (x_0 + \xi (s_n))\Big] \label{step1} \\
& = & -i\frac{d \xi_i (s')}{d s'}A_i (x_0 + \xi (s')) \star_{x_0} \Big[1 + \sum^{\infty}_{k=1} (-i)^{k} \ \int^{s'}_0 \! d s_2
\cdots \int^{s_{k}}_0 \! d s_{k+1}  \ \nonumber \\
& & \times \frac{d \xi_{i_2} (s_2)}{d s_2} \cdots \frac{d \xi_{i_{k+1}} (s_{k+1})}{d s_{k+1}}
A_{i_2} (x_0 + \xi (s_2)) \star_{x_0} \cdots \star_{x_0} A_{i_{k+1}} (x_0 + \xi (s_{k+1})) \Big] \hspace{30pt} \label{step2} \\
& = & -i\frac{d \xi_i (s')}{d s'}A_i (x_0 + \xi (s')) \star_{x_0} \Big[1 + \sum^{\infty}_{k=1} (-i)^{k} \ \int^{s'}_0 \! d s_1
\cdots \int^{s_{k-1}}_0 \! d s_{k}  \ \nonumber \\
& & \times \frac{d \xi_{i_1} (s_1)}{d s_1} \cdots \frac{d \xi_{i_{k}} (s_{k})}{d s_{k}}
A_{i_1} (x_0 + \xi (s_1)) \star_{x_0} \cdots \star_{x_0} A_{i_{k}} (x_0 + \xi (s_{k})) \Big] \hspace{30pt} \label{step3} \\
& = & -i\frac{d \xi_i (s')}{d s'}A_i (x_0 + \xi (s')) \star_{x_0} \mathcal{U} (x(s'), x_0, C) \label {step4},
\end{eqnarray}
where the names of the dummy indices of summation have been renamed to
$n-1 = k$ and $i_1 = i $ in going from
equation (A.\ref{step1}) to equation (A.\ref{step2}). In equation (A.\ref{step3}), the integration variables
have been renamed from equation (A.\ref{step2}) by decrementing the value of k by 1 in order to make the result more transparent.
Note that this calculation could be done because the
star-product is evaluated between $x_0$:s and does not influence the integration.

The newly obtained relation (A.\ref{step4}) can also be written in the form
\begin{equation}
\frac{d \xi_i (s')}{d s'}{d \over d\xi_i(s')} \mathcal{U} (x(s'), x_0, C) =
-i\frac{d \xi_i (s')}{d s'}A_i (x_0 + \xi (s')) \star_{x_0} \mathcal{U} (x(s'), x_0, C).
\label{rela2}
\end{equation}
If we then go back to the path ordered exponential as given by (A.\ref{po-exp2}) and consider
it as a function depending on two points, the initial and final point, we
notice that we can interpret $\xi_i(s')$ as the point $l_i = \xi_i(1)$ in the
parametrization of (A.\ref{po-exp2}). This leads to the relation
\begin{equation}
{d \over dl_i} \mathcal{U} (x, x_0, C) =
-iA_i(x_0 + l) \star_{x_0} \mathcal{U} (x, x_0, C),
\label{rela1}
\end{equation}
from equation (A.\ref{rela2}).
This relation can be written in the form (A.\ref{wilrel}) by noting that since $x_i={x_0}_i + l_i$ we have
relations of the form
\begin{eqnarray}
{d \over dl_i} & = & {d({x_0}_i + l_i)\over dl_i}{d\over d({x_0}_i + l_i)} = {d\over d({x_0}_i + l_i)} = {d\over dx_i} \\
{d \over d{x_0}_i} & = & {d({x_0}_i + l_i)\over d{x_0}_i}{d\over d({x_0}_i + l_i)} = {d\over d({x_0}_i + l_i)} = {d\over dx_i},
\end{eqnarray}
because $x, x_0$ and $l$ must be independent variables for the NC path ordered
exponential (A.\ref{po-exp2}) to be sensibly defined. As a result
\begin{equation}
\Big({d \over d{x_0}_i}\Big)^n = \Big({d\over dx_i}\Big)^n,
\end{equation}
so that the star product with respect to $x_0$ in (A.\ref{rela1}) can safely be transformed into a star
product with respect to $x$ and therefore we finally have
\begin{equation}
{d\over dx_i} \mathcal{U} (x, x_0, C) = -iA_i(x)\star_x\mathcal{U}(x, x_0, C),
\end{equation}
which is exactly (A.\ref{wilrel}) or (\ref{Ansatz2}).

%%%%%%%%%%%%%%%%%%%%%%%%%%%%%%%%%%%%%%%%%%%%%%%%%%%%%%%%%%%%%%%%%%%%%%%%%%%%%%%%%%%%%%%%%%

%%%%%%%%%%%%%%%%%%%%%%%%%%%%%%%%%%%%%%%%%%%%%%%%%%%%%%%%%%%%%%%%%%%%%%%%%%%%%%%%%%%%%%%%%%

\renewcommand{\baselinestretch}{1}\selectfont

\end{document}